\documentclass{aastex}          %% The manuscript based on AASTeX v5.x
\usepackage{spr-astr-addons}    %% mimicing ASTR journal style
\begin{document}
%% Article title
%
\title{The Mystery of Unidentified Infrared Emission Bands}

%% Running heads
\shorttitle{}
\shortauthors{<Kwok>}

%% Author and Affilations
\author{Sun Kwok\altaffilmark{1}}\email{skwok@eoas.ubc.ca} %% non-output

\altaffiltext{1}{Department of Earth, Ocean, and Atmospheric Sciences, University of British Columbia, Vancouver, Canada; corresponding author:skwok@eoas.ubc.ca}

%\noindent{Email address of corresponding author Sun Kwok: skwok@eoas.ubc.ca}\\
\noindent{Astrophysics and Space Science (in press)\\

%% Abstract
\begin{abstract}
A family of unidentified infrared emission (UIE) bands has been observed throughout the Universe.  
The current observed spectral properties of the UIE bands are summarized.  These properties are discussed in the frameworks of different models of the chemical carriers of these bands.  The UIE carriers represent a large reservoir of carbon in the Universe, and play a significant role in the physical and chemical processes in the interstellar medium and galactic environment.  
A correct identification of the carrier of the UIE bands is needed to use these bands as probes of galactic evolution.

\end{abstract}

\keywords{astrobiology; astrochemistry; ISM: lines and bands; ISM: molecules;  planetary nebulae: general; reflection nebulae; stars: AGB and post-AGB; galaxies: starburst}

\section{Discovery of the unidentified infrared emission bands}

When astronomical infrared observational capabilities first developed in the 1960s, the expectation was that the infrared spectra of emission nebulae would  be dominated by atomic fine-structure lines.  The detection of a strong infrared excess in the planetary nebula NGC 7027 was totally unexpected \citep{gillett1967}.  The large infrared continuum emission in NGC 7027 and other planetary nebulae was later identified as thermal emission from circumstellar solid-state grains.  Even more surprising was that on top of this strong infrared continuum, there is a strong emission feature at 11.3 $\mu$m that cannot be identified with any known atomic lines  \citep{gillett1973}.  
Further ground-based observations found another strong unidentified emission feature at  3.3 $\mu$m in NGC 7027  \citep{merrill1975}. 
\citet{russell1977a} found that the strength of the 11.3\,$\mu$m feature closely correlates with the 3.3\,$\mu$m feature, suggesting a common origin for the two features. 
Since the 3.3 and 11.3 $\mu$m features are too broad to be atomic lines and show no substructures to qualify as molecular bands, they were believed to arise from mineral solids such as carbonates \citep{gillett1973, russell1977a}.  This seemed a reasonable interpretation at the time because silicate minerals have already been found to be common in the circumstellar envelopes of  evolved stars \citep{woolf1969}.

As ground-based observations in the infrared are limited by the availability of atmospheric windows, there was a spectral gap between 4--8 $\mu$m that remained unexplored.  %further exploration of the spectra of planetary nebulae have to relied on space-based observations.   
From  spectrophotometric observations obtained from  the {\it Kuiper Airborne Observatory (KAO)}, additional features at  6.2, 7.7, and 8.6 $\mu$m were discovered  \citep{russell1977b, russell1978}.  The failure to detect the expected 7.0 $\mu$m carbonate feature ruled out the mineral hypothesis.   Since the 3.3, 6.2, 7.7, 8.6, and 11.3 $\mu$m features have no counterparts in the infrared spectra of atoms, they are collectively known as the unidentified infrared emission (UIE) bands.

The UIE bands are found to be present in planetary nebulae, reflection nebulae, H{\sc ii} regions, novae, as well as spatially widely distributed in the diffuse interstellar  medium in the Milky Way Galaxy and in external galaxies.  

In addition to their ubiquitous nature, the most intriguing element of the UIE phenomenon is that it is a manifestation of organic matter in space.  Among the first to suggest that there could be an organic component in the interstellar medium was Bertram Donn, who proposed polycyclic hydrocarbons as a possible component of interstellar grains \citep{donn1968}.  After the discovery of the UIE bands,  \citet{hoyle1977} suggested that the spectrum of polysaccharides shows resemblance to the astronomical UIE bands.  

The infrared spectra of organic compounds have actually been well studied by chemists since the 1960s.  This connection between astronomical spectra and laboratory spectroscopy was made by \citet{knacke1977}, who identified the 3.3 $\mu$m UIE band as originating from C$-$H stretching mode of aromatic compounds.  The possible organic origin of the UIE feature were also raised by  \citet{puetter1979}.
\citet{sagan1979} suggested a connection between the astronomical UIE bands and the laboratory-synthesized complex organic polymer tholins \citep{khare1973}.  The organic origin of the UIE bands was extensively discussed by \citet{duley1981}, who assigned the 3.3 and 11.3 $\mu$m features to  aromatic materials. All these were treated as speculations at the time and  received little attention from the astronomical community.  %The organic hypothesis was only taken seriously after advances in microwave technology led to the detection of numerous interstellar molecules through their rotational transitions.

\section{The UIE phenomenon}

\begin{figure}
\begin{center}
\includegraphics[width=\columnwidth]{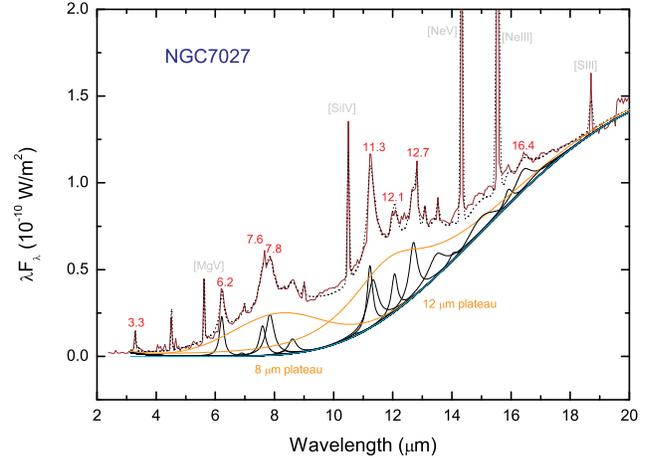}
\end{center}
\caption{UIE  bands in the planetary nebula NGC 7027 as observed by the {\it Infrared Space Observatory}
.  The observed spectrum is spectrally decomposed into the UIE bands (in  black), plateau features (in orange), and the underlying dust continuum (in blue).  The wavelengths of the UIE bands are labeled in red.  The narrow features (labeled in grey) are atomic lines.  Figure adapted from  \citet{kz2011}.}
\label{ngc7027}
\end{figure}

Although most of the discussions of UIE bands have focused on the strong  3.3, 6.2, 7.7, 8.6, and 11.3 $\mu$m bands, the UIE phenomenon is much richer than these features.  
%Also present in astronomical spectra are emission features around 3.4 $\mu$m,   which arise from symmetric and asymmetric C--H stretching modes of methyl and methylene groups \citep{geballe1992}.  The bending modes of these groups also manifest themselves at 6.9 and 7.3 $\mu$m \citep{jou90,chiar2000, kwo01}. 
There are a number of minor emission features at  12.1, 12.4, 12.7, 13.3, 15.8, 16.4, 17.4, 17.8, and 18.9 $\mu$m, which have been observed in proto-planetary nebulae \citep{kwok1999}, reflection nebulae \citep{sel07}, and galaxies \citep{stu00} (Figures \ref{ngc7027} and  \ref{ngc7023}).
%Sellgren 2007: 15.9, 16.4, 17.0, 17.4, 17.8, and 18.9 um
%Strum 13.6, 15.8, and 16.5 um
%unassigned features at 15.8, 16.4, 17.4, 17.8 and  18.9\,$\mu$m \citep{jd90, kv99, ct00, sturm2000, sellgren2007}. 
Most significantly, the UIE features are often accompanied by strong, broad emission plateaus features at 6--9, 10--15, and 15--20 $\mu$m (Figs.~\ref{ngc7027} and \ref{ngc7023}).  

\begin{figure}
\begin{center}
\includegraphics[width=\columnwidth]{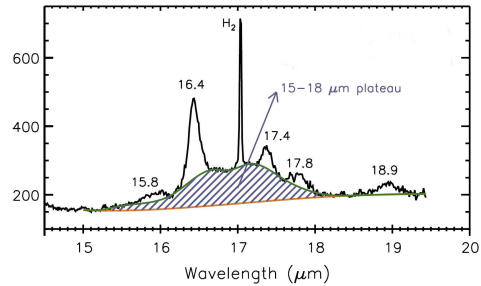}
\end{center}
\caption{{\it Spitzer} IRS spectrum of NGC 7023 showing the 15--20 $\mu$m broad emission plateau (the shaded region) centered around 17 $\mu$m and the minor UIE bands at 15.8, 16.4, 17.4, and 17.8 $\mu$m.  The 18.9 $\mu$m feature is assigned to C$_{60}$.  Figure adapted from  \citet{peeters2012}.}
\label{ngc7023}
\end{figure}

\subsection{Discovery of aliphatic organics}

%Although the 3.3 and 11.3 $\mu$m features have been attributed to aromatic materials \citep{duley1981}, the detection of 3.4 $\mu$m aliphatic feature in sources showing 3.3 $\mu$m feature suggests that the carriers of UIE bands are not purely aromatic.
Motivated by Hoyle's idea that organic matter is common in interstellar space, \citet{wick1980} undertook a search for absorption signatures along the sight to the Galactic Center and detected a strong signal at 3.4 $\mu$m.  This feature is most likely due to C$-$H stretch of aliphatic compounds, which \citet{duley1979} had suggested to be observable through absorption spectroscopy toward infrared sources.
This work was followed by a number of studies using both ground-based and space-based telescopes \citep{pendleton1994, sandford1991, chiar2000, dartois2004a}, leading to the discovery of the aliphatic C$-$H bending mode at 6.9 $\mu$m.  The 3.4 $\mu$m feature has also been detected in external galaxies \citep{imanish2000, imanishi2010}
The strength of the 3.4 $\mu$m feature suggests that at least 15\% of all the carbon atoms are in the form of aliphatic compounds \citep{dartois2011}

The same 3.4 $\mu$m feature was also found in emission in planetary nebulae showing the 3.3 $\mu$m feature \citep{Jourdain86}, which implies that the carriers of UIE bands are not purely aromatic. Although the 3.3 $\mu$m band is generally much stronger than the 3.4 $\mu$m band, in some sources (e.g., proto-planetary nebulae) the two features can be comparable in strength \citep{geballe1992, goto2007}.  Distinct components of the 3.4 $\mu$m feature at 3.40, 3.46, 3.52, and 3.56 $\mu$m have been attributed to symmetric and anti-symmetric C$-$H stretching modes  \citep{deMuizon1990, hrivnak2007}.

\subsection{Band positions and profiles}

The observed astronomical UIE spectra can be classified into different groups based on the band positions and profiles.  \citet{peeters2002} and \citet{van04} classified the spectra into classes \textsl{A}, \textsl{B}, \textsl{C}, and \textsl{D}, with the 7.7 $\mu$m feature showing the largest variation among the classes.  The 7.7 $\mu$m band shows peak positions at 7.6 and 7.8 $\mu$m, and sometimes at 8.0, and 8.2 $\mu$m, and these variations are used to designate the different classes (right panel, Figure \ref{peeters}).  The 6.2 $\mu$m feature sometimes peaks at wavelength as long as 6.29 $\mu$m (left panel, Figure \ref{peeters}).  Generally speaking, H{\sc ii} regions, reflection nebulae, and galaxies belong to Class \textsl{A}, Herbig AeBe stars to class \textsl{B}, and post-Asymptotic-Giant-Branch stars to Class \textsl{C}.  
This classification scheme is also applied to UIE sources in the Magellanic Clouds \citep{ sloan2014}.

\begin{figure}
\begin{center}
\includegraphics[width=\columnwidth]{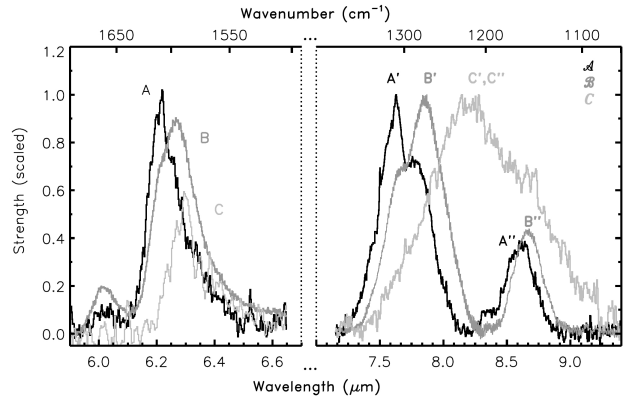}
\end{center}
\caption{The band position and profiles of the 6.2 $\mu$m (left panel) and 7.7/8.6 $\mu$m (right panel) features in different classes of objects.  Class \textsl{A}  objects have band peaks at 6.2 (A), 7.6 (A$^\prime$) and 8.6 $\mu$m (A$^{\prime\prime}$), Class \textsl{B} objects have band peaks at 6.25 (B), 7.8 (B$^\prime$), and 8.8 $\mu$m (B$^{\prime\prime}$).  Class \textsl{C} objects have band peaks at 6.3 (C), broad 8.2 $\mu$m (C$^\prime$ and C$^{\prime\prime}$). Figure adapted from  \citet{peeters2002}.}
\label{peeters}
\end{figure}

\subsection{Underlying continuum}
\label{continuum}
The UIE bands are often observed in emission on top of a continuum (Figures \ref{ngc7027} and \ref{m82}).  This continuum is attributed to thermal emission from micron-size solid particles (``dust'' in astronomical nomenclature).  Since this continuum is often featureless, it is assumed to originate from amorphous carbon grains.  In the diffuse interstellar medium, the strengths of the UIE bands are correlated with the strength of the dust continuum \citep{kahanpaa2003}.  Such correlations imply that the UIE bands and the dust continuum must share the same heating source.

\begin{figure}
\begin{center}
\includegraphics[width=\columnwidth]{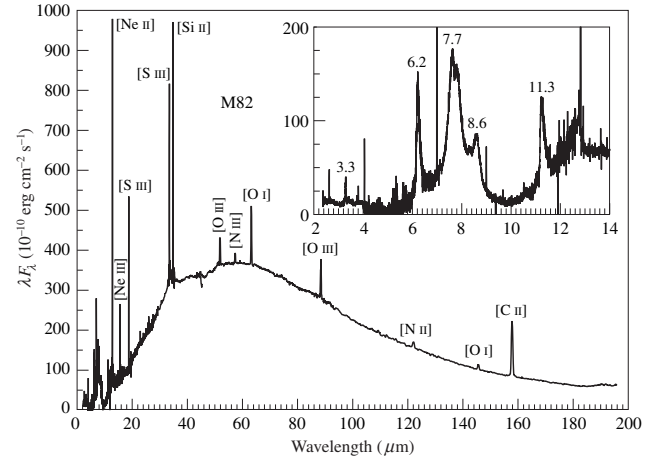}
\end{center}
\caption{The UIE bands (expanded in insert box) of the starburst galaxy M82 are observed on top of an infrared continuum with a color temperature of $\sim$50 K.  The narrow lines are atomic lines.  Figure adapted from \citet{kwok2007}}
\label{m82}
\end{figure}

\subsection{Association with fullerenes}
Since the initial discovery of fullerene (C$_{60}$) in the planetary nebula Tc-1 \citep{cami2010}, C$_{60}$ emissions have been detected in reflection nebulae \citep{sellgren2010}, planetary nebulae \citep{garcia2010}, and proto-planetary nebulae \citep{zhang2011} where UIE bands are simultaneously present.  Although Tc-1 has no strong UIE features, it does show the broad 8 and 12 $\mu$m plateaus features. 
% has generated interests in possible links between C$_{60}$ and the UIE phenomenon.  (Fig.~\ref{tc1}). The proto-planetary nebula IRAS 01005+7910, however, show C$_{60}$ features, UIE bands, as well as the 8, 12, and 17 $\mu$m emission plateau features .   
The association between C$_{60}$ and the 8 and 12 $\mu$m plateau features \citep{zhang2013, otsuka2013} suggests that amorphous carbonaceous solids could be precursors of fullerenes \citep{garcia2012, bernard2012}.

\section{Chemical nature of the carriers}\label{carriers}
Although the carrier of the UIE bands are generally recognized as due to a carbonaceous compound since the work of \citet{duley1981}, the exact chemical composition and structure of the carrier is not yet settled.  Some of the models are discussed below.
%A summary of proposed carriers can be found in \citet{yang2017}

\subsection{Polycyclic aromatic hydrocarbon molecules}

The idea that UIE bands originate from polycyclic aromatic hydrocarbon (PAH) molecules was proposed by  \citet{leger1984} and \citet{allamandola1985}.  \citet{allamandola1985} calculated the vibrational spectrum of gas-phase chrysene and suggest possible matches to the UIE bands. The PAH hypothesis was developed based on (i) some general similarities between the infrared spectra of UIE bands with PAH molecules; and (ii) the 12 $\mu$m excess emission observed in cirrus clouds in the diffuse interstellar medium can be explained by single-photon excitation of small molecules  \citep{sellgren1984, sellgren2001}.   The physics of PAH molecules as a component of the interstellar medium was discussed by \citet{omont1986}.

The thesis of the PAH hypothesis is summarized by  \citet{tielens2008} as ``These features are (almost) universally attributed to the IR fluorescence of far-ultraviolet (FUV)-pumped polycyclic aromatic hydrocarbon (PAH) molecules, containing ~50 C atoms''. 
The small size of the PAH molecules allow them to be stochastically excited to temperature of 1,000 K upon absorption of a single UV photon, and subsequent cascade produces the vibrational emission bands in the infrared.   
In the past 30 years, the PAH hypothesis has become extremely popular and is commonly accepted in the astronomical community as the explanation of the UIE phenomenon. 
The 3.3 $\mu$m UIE band is assigned to C$-$H stretching modes, the 6.2 $\mu$m and 7.7 $\mu$m features to C$-$C stretching modes, the 8.6 $\mu$m feature to C$-$H in-plane bending modes, and the 11.3 $\mu$m to C$-$H out-of-plane bending modes of PAH molecules \citep{all89}.  Details of the PAH hypothesis are reviewed by \citet{peeters2011} and \citet{ peeters2021}.  The application of PAH hypothesis to the study of galaxies is reviewed by \citet{li2020}.  The spectral properties of PAH molecules have been extensively studied \citep{hudgins2004, salama2008} and collected into a comprehensive database \citep{bb10, bb14, mattioda2020}.

The first serious objections to the PAH hypothesis was given by \citet{donn1989}.  Below is a list of problems with the PAH hypothesis, expanded and updated from the list of \citet{donn1989}.  

\begin{itemize}
  \item {PAH molecules have well-defined sharp absorption bands but the UIE bands are broad (Figs.~\ref{ngc7027} and \ref{peeters}).  Superpositions of vibrational bands from a large variety of PAH molecules  and artificial broadening profiles have to be introduced  to fit the astronomical spectra \citep{li2001, dra07, cami2011}.}
  \item Neutral PAH molecules are primarily excited by UV and have little absorption in the visible \citep{ciar1964, leger1989}, 
 but UIE features are seen in proto-planetary nebulae and reflection nebulae with no background UV radiation.
  \item The expected strong absorption features of PAH molecules in the UV are not seen in interstellar extinction curves. The observed PAH to hydrgogen abundance ratio upper limits range from $10^{-10}$ to $10^{-8}$ \citep{clayton2003, salama2011, gredel2011}, which are much lower than the PAH/H ratio of $3\times10^{-7}$ predicted from the strength of the infrared features \citep{tielens2008}.
  \item The infrared spectra of small PAH molecules are well studied by chemists \citep{schlemmer1994, cook1996, cook1998} and they found ``No PAH emission spectrum has been able to reproduce the UIE spectrum w.r.t. either band positions or relative intensities'' \citep{ wagner2000}.
  \item The PAH model predicts that the UIE band ratios are strongly dependent on the UV background as individual UIE bands are assumed to arise from different neutral or ionized PAHs \citep{hudgins2004, draine2021}.  But the shapes and peak wavelengths of UIE bands are found to be independent of temperature of exciting star, ranging from thousands to tens-of-thousands of degrees \citep{uchida2000} (Fig.~\ref{uchida}). An analysis of 820 UIE spectra in the diffuse interstellar medium shows no variation of the UIE band ratios in regions of widely different UV backgrounds \citep{chan2001}.
   \item {Presence of aliphatic features in UIE sources.  In the PAH hypothesis, the 3.4 $\mu$m feature is interpreted as superhydrogenation of PAH molecules \citep{sch93}. 
It has also been suggested as arising  from hot bands from anharmonic aromatic C$-$H stretch, which shifts the 3.3 $\mu$m feature to a longer wavelength  \citep{barker}.   However, theoretical calculations including anharmonicity show a simultaneous increase of the width of the 3.3\,$\mu$m feature, which was not observed \citep{van04}. Furthermore, the expected strong overtone bands at the 1.6--1.8\,$\mu$m region is not detected, making the anharmonicity explanation unlikely \citep{Goto2003}. }
  
\end{itemize}

\begin{figure}
\begin{center}
\includegraphics[width=\columnwidth]{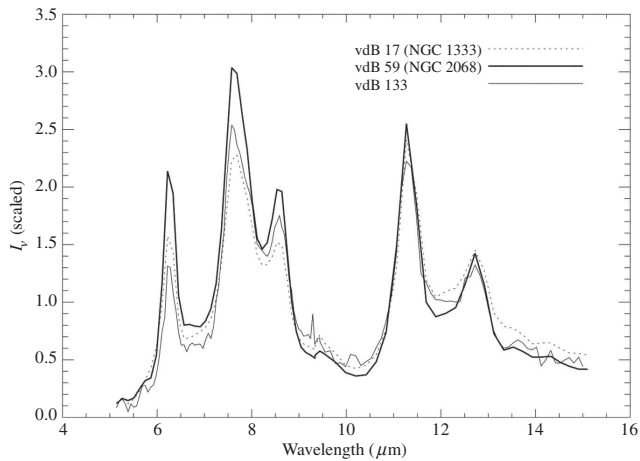}
\end{center}
\caption{Normalized {\it Infrared Space Observatory} spectra of three reflection nebulae showing that  the UIE bands have similar peak wavelengths, spectral shapes, and
continuum levels independent of the temperatures of the illuminating stars (vdB 17, 11,000 K; vdB 59, 19,000 K; vdB 133 6,800 K). Figure adapted from \citet{uchida2000}.}
\label{uchida}
\end{figure}

An independent way to establish the presence of PAH molecules in space is to search for them through their rotational transitions.  However, due to the absence of permanent dipole moment of these molecules, this is not as feasible as for other classes of molecules.   The simplest ring molecule benzene is detected via its infrared transitions in the proto-planetary nebula AFGL 618 \citep{cernicharo2001}.
%, showing that molecular synthesis is active in the late stages of stellar evolution.  
Two double-ring molecules with nitrile functional groups (1- and 2-cyanonaphthalene, $c-$C$_6$H$_5$CN) have been detected in the molecular cloud TMC-1 \citep{mcguire2021}. Two pure hydrocarbon cyclic molecules:  cyclopentadiene ($c-$C$_5$H$_6$, a 5-ring molecule) and indene($c-$C$_9$H$_8$, a bicyclic molecule with both a 5- and 6-membered ring) have been detected in TMC-1  \citep{cernicharo2021, burkhardt2021}.

In part to address the above list of problems, the PAH hypothesis has been revised to incorporate different ionization states and large sizes to increase the absorption cross sections in the visible.  The molecular size range has been extended to $>$1,000 atoms \citep{peeters2021}.
The PAH molecules are also modified to include  dehydrogenation, superhydrogenation and minor aliphatic side groups \citep{li2012}.
%Anharmonic shifts and hot bands
Heteroatoms (defined as those atoms other than C or H) such as N  are also introduced to explain the 6.2 $\mu$m band \citep{hudgins1999}.

In order to fit the astronomical observations, the PAH model appeals to a mixture of PAH of different sizes, structures (compact, linear, branched) and ionization states, as well as artificial broad intrinsic line profiles \citep{all99, cami2011}.  The number of free parameters is so large that  the routines used by the PAH model to fit the astronomical UIE spectra have been shown to be able to fit any artificial spectra \citep{zhang2015}.

\subsection{Other hydrocarbons}

In addition to pure carbon materials such as graphite (hybridization state $sp^2$) and diamond (hybridization state $sp^3$), different amorphous C$-$H alloys can be created when H is introduced \citep{cataldo2004,jones2013}. 
These amorphous solids consist of varying degrees of  aromatic to aliphatic ratios, number of aromatic rings, different lengths of aliphatic chains, and arranged in different geometric structures.  A schematic of possible structures of amorphous carbonaceous solids is shown in Figure \ref{robertson}.  The lower left corner of the triangle represents graphite (C rings on a plane with no H), the top corner represents diamonds (C arranged in tetrahedral forms), PAHs (aromatic rings arranged on a plane) are on the bottom edge, and various forms of amorphous hydrogenated carbon can exist in the interior of the triangle.  Theoretical calculation on the infrared spectra of such amorphous solids suggest that they can  be good candidates as carrier of UIE bands \citep{jones2012a, jones2012b}.  
%The infrared spectra of these amorphous carbonaceous materials \citep{dischler1983} resemble the astronomical UIE bands seen in planetary nebulae and proto-planetary nebulae (Figure \ref{dischler}).  
Since these amorphous carbonaceous solids have absorption bands in the visible, they can be easily excited by visible light from stars.

\begin{figure}
\begin{center}
\includegraphics[width=\columnwidth]{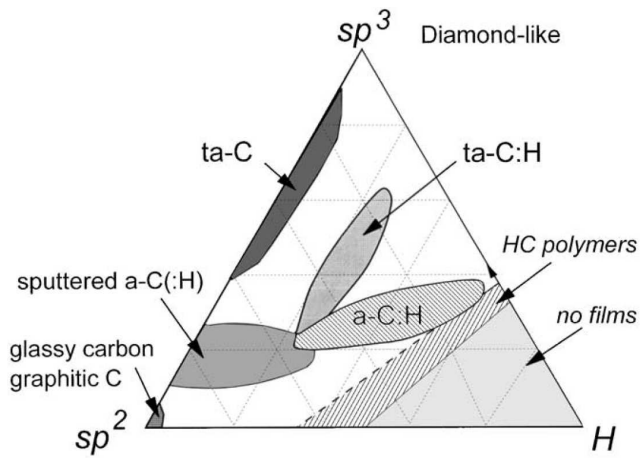}
\end{center}
\caption{Phase diagram of amorphous carbonaceous (pure C and H) compounds.  The lower right corner of the triangle represents pure H, lower left corner pure graphite-like ($sp^2$) materials, and the upper corner pure diamond-like ($sp^3$).  Areas inside the triangle represent various H/C ratios and $sp^2/sp^3$ mixed hybridization states.  Figure adapted from  \citet{robertson2002}.}
\label{robertson}
\end{figure}

The possibility that the UIE bands could be due to some form of hydrogenated amorphous carbon (HAC) was made by  \citet{buss1990}, and HAC was proposed to be a major constituent of interstellar dust by \citet{jones1990}.  Further work on HAC was described in  \citet{duley1993} and \citet{duley2000}.  The infrared properties of HAC were studied in the laboratory by \citet{scott1997}, \citet{duley2005}, and \citet{gad2012}.  

By subjecting methane gas in a vacuum to microwave radiation, \citet{sakata1983} were able to collect carbonaceous composite particles  on a substrate after rapid cooling, which they name quenched carbonaceous composite  (QCC).  The infrared spectra of QCC show strong resemblance to the astronomical UIE bands  \citep{sakata1984, sakata1987}.

%carbonaceous molecules \citep{bernstein2009},  

By accreting C, C$_2$, and other C$_n$ molecules, carbon nanoparticles with sizes less than 100 nm can be created in the laboratory. The infrared spectra of these particles show similarities to the astronomical UIE spectra  \citep{hu2006, hu2008}.

\subsection{Amorphous hydrocarbons with heteroatoms}

Mixed aromatic/aliphatic structures are natural products of combustion. The first nucleation products of soot particles formed in flames have structures consisting of islands of aromatic rings linked by chains \citep{col1997, chung2011}.  
In natural interstellar or circumstellar environments, other elements such as O, N, and S are abundantly present and can be expected to be incorporated into any carbonaceous compounds condensed from gas phase.  
Based on the similarities between the astronomical UIE spectra and the infrared spectra of coal,  \citet{papoular1989} suggested that coal as the carriers of the UIE bands.   This idea was later extended to kerogen-like materials \citep{papoular2001}.  Petroleum fractions as a carrier of UIE bands are discussed by \citet{cataldo2002}.  Coal, kerogen, and petroleom are natural products that contain heteroatoms in their structures.  The model of  mixed aromatic/aliphatic organic nanoparticles (MAON) considers a carbonaceous compound containing aromatic rings of different sizes and aliphatic chains of different lengths and orientations arranged in a 3-D amorphous structure, mixed with heavy elements  \citep{kz2013}.
Fig.~\ref{maon} shows a schematic structure of MAON, illustrating the complexity of the chemical structure, as well as the kinds of functional groups (including heteroatoms) that may be contained in such structures.  %However, the direct detection of such impurities would require higher sensitivity and spectral resolution than our existing  infrared spectroscopic facilities can offer.

\begin{figure}
\begin{center}
\includegraphics[width=\columnwidth]{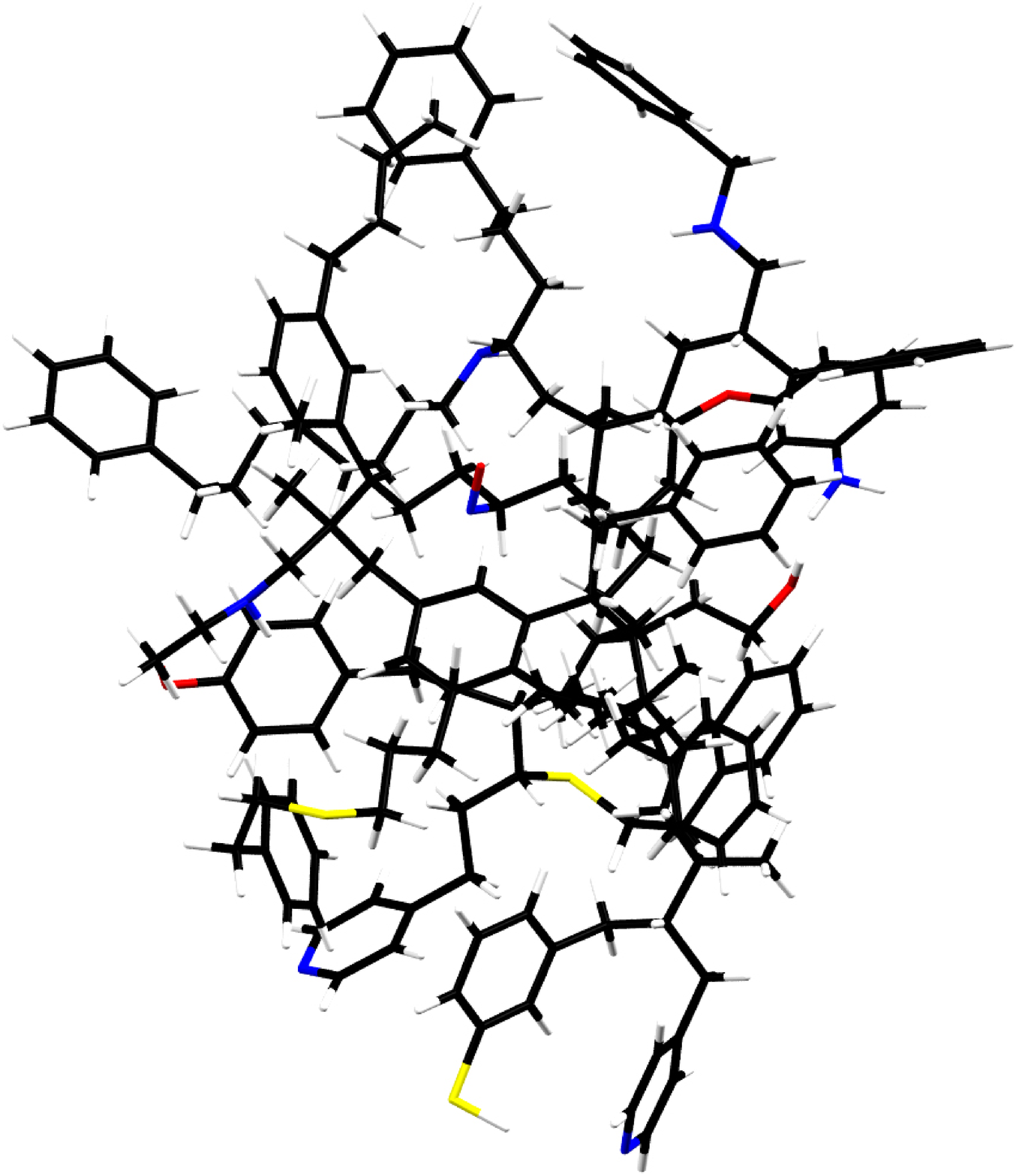}
\end{center}
\caption{The MAON structure is characterized by a highly disorganized arrangement of small units of aromatic rings linked by aliphatic chains.  This structure contains 169 C atoms (in black) and  225 H atoms (in white).  Impurities such as O (in red), N (in blue), and S (in yellow) are also present.  A typical MAON particle may consist of multiple structures similar to this one.}
\label{maon}
\end{figure}

By introducing nitrogen gas into the QCC experiment, a solid condensate quenched nitrogen-included carbonaceous composite (QNCC) is collected.  The infrared spectra of QNCC show resemblance of the UIE spectra of novae \citep{endo2021}.

\section{Other synthesized substances}

Since the early work on tholins \citep{khare1973},  various carbonaceous compounds have been synthesized in the laboratory by subjecting simple hydrocarbons to different forms of energy injections.
%laboratory often results in disorganized materials with mixed aromatic/aliphatic structures.  
The techniques used  include laser ablation of graphite \citep{scott1996, mennella1999, jager2008}, laser pyrolysis of gases \citep{herlin1998}, arc discharge \citep{mennella2003},  microwave irradiation \citep{sakata1984, wada2009, god2011}, UV photolysis \citep{dartois2004b}, and flame synthesis \citep{car2012}. 
Chemical analysis of these laboratory synthesized carbonaceous nanoparticles show that they consist of networks of chains and rings \citep{hu2006} and their spectra show varying degrees of resemblance to the astronomical UIE spectra.

\section{Atomic origin}
The first proposal for an atomic origin of the UIE bands was made by \citet{holmlid2000} who suggests that the astronomical bands are the result of electronic deexcitation of Rydberg matter states with principal quantum numbers $n=40-200$.   
Rydberg matter is proposed to be a new phase of matter equivalent to that of liquid and solid which could exist in space in large quantities.  
In support of this hypothesis,  \citet{holmlid2000} shows that the Rydberg matter model can fit the observed UIE spectra with fewer parameters than the PAH model.
This idea was extended by  \citet{zagury2021} who assigns the 3.3, 6.2, and 11.3 $\mu$m UIE bands to hydrogen recombination line series $n$=6, 8, and 11, respectively, and the 15$-$20 $\mu$m plateau emission band to the continua of hydrogen transitions to $n=13$ and $n=14$.  
However, the hydrogen hypothesis is contradicted by the fact that the UIE phenomenon is observed only in carbon-rich sources \citep{cohen1986, cohen1989}.

\section{Corresponding vibrational modes of the UIE features}

Since the UIE phenomenon consists of a family of emission features, a successful theory must be able to consistently explain the positions, profiles, relative strengths of all the entire family.  The ideal candidate for the carrier of UIE bands should consistently produce the entire family of UIE bands without appealing to special conditions.
PAH molecules show spectral features near 3.3 and 11.3 $\mu$m but do not have clear counterparts for the 6.2, 7.7, and 8.6 $\mu$m UIE bands.
In the words of \citet{cook1998}: ``In order to reproduce the narrow 6.2 and 11.2 $\mu$m UIR bands, the carriers must consistently exhibit bands at these positions with a consistency similar to that which is observed with the 3.3 $\mu$m emission. In addition, the carriers of the UIRs must, in general, exhibit an absence of strong bands in the gap between the 6.2 and 7.7 $\mu$m UIR features. The PAHs used in these model spectra simply do not meet these criteria; hence they do not reproduce the details of the UIR spectra.''

One of the advantages of the HAC/MAON-type of models is that they naturally produce broad infrared features without appealing to artificial fittings \citep{dischler1983, guillois1996}.  The exact positions of the bands depend on the composition of the material (C to H ratio, fraction of impurities), aromatic to aliphatic ratios, and geometric factors, but their resemblance to the astronomical UIE bands is unmistakable \citep[Fig.~9,][]{kwok2016}.  However, there is no specific vibrational mode assignments to these observed bands and the molecular dynamical origin of the strong bands observed in laboratory specimens   \citep[e.g., those observed in][]{dischler1983, herlin1998} are not identified.
In order to advance the MAON hypothesis further, we need to perform quantum chemistry calculations of large varieties of MAONs to identify the major vibrational modes as well as to see whether some of these modes will converge to resemble the observed astronomical UIE bands 
 \citep{sadjadi2016}.

\subsection{The 3.3 $\mu$m band}

The 3.3 $\mu$m UIE feature was first identified as aromatic C$-$H stretch by \citet{knacke1977}.  This is a prominent band and shows a consistent emission profile in many astronomical sources (Figure \ref{togunaka}).  The measured central wavelength of the feature is 3.2887$\pm$0.0009 $\mu$m  \citep{tokunaga2021}.  
The invariance in feature profile, independent of source excitation conditions and evolutionary history, puts severe constraints on the chemical composition of the carrier and excitation mechanism.
The central wavelengths of the C$-$H stretching mode of PAH molecules lie shortward of the observed wavelength of the 3.3 $\mu$m band \citep{sakata1990, kz2013}.  %If the 3.3 $\mu$m feature arises from PAH molecules, then it cannot be due to neutral species. 
 \citet{joblin1995} consider the effects of anharmonic couplings and suggest that the astronomical 3.3 $\mu$m feature can be consistent with PAH molecules at high temperatures.  But the observed peak wavelengths of the 3.3 $\mu$m feature lie within a very narrow range, even when observed in circumstellar or interstellar environments under very different temperature conditions.  This makes it unlikely that red shift by high temperatures be the general cause of this wavelength discrepancy.

\begin{figure}
\begin{center}
\includegraphics[width=\columnwidth]{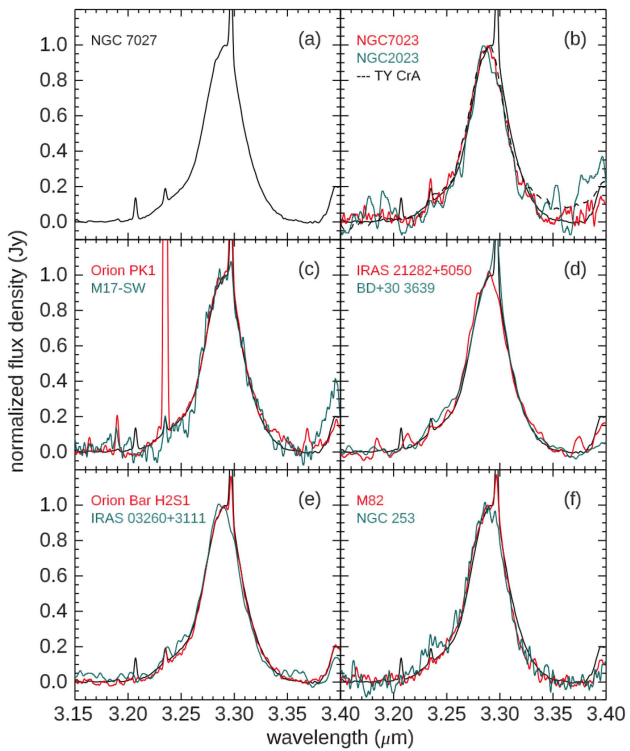}
\end{center}
\caption{Profiles of the 3.3 $\mu$m feature observed in NGC 7027 (plotted in black in all panels) and other astronomical sources (plotted in red and green in panels b-f, as well as in black dashed line in panel b).  The sharp lines at 3.207, 3.234, and 3.297 $\mu$m are the
[Ca IV], H$_2$ 1$-$0 O(5), and H Pfund-$\delta$ lines, respectively.  Figure adapted from  \citet{tokunaga2021}.}
\label{togunaka}
\end{figure}

\begin{figure}
\begin{center}
\includegraphics[width=\columnwidth]{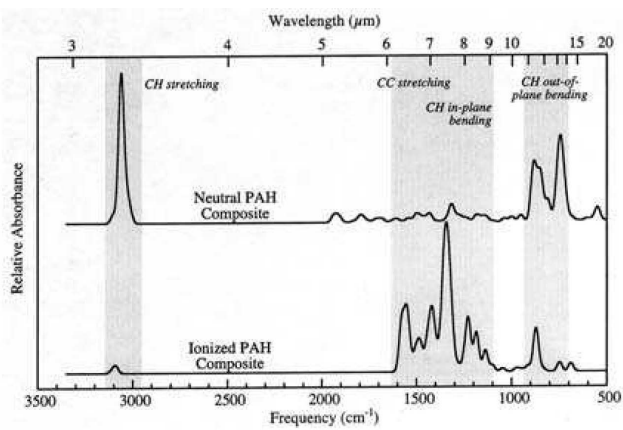}
\end{center}
\caption{A comparison of the absorption band strengths between neutral (top curve) and ionized (bottom curve) PAH molecules.  The UIE bands in the 6$-$9 $\mu$m region are attributed to ionized PAH molecules in the PAH hypothesis.   Figure adapted from \citet{hudgins2004}.}
\label{hudgins}
\end{figure}

Ground-based high-spectral resolution observations of the 3.3 $\mu$m of HD44179 (the Red Rectangle) suggest that the 3.3 $\mu$m feature can be decomposed into two components, one at 3.28 $\mu$m and another at 3.30 $\mu$m \citep{Song03}.
The existence of these two components has been attributed to 
%One is based on the PAH model where the 2.28 and the 3.30 $\mu$m components arise from 
``bay'' and ``non-bay'' hydrogen sites of the PAH units, based on 
%This interpretation is  mainly based on 
the detection of two bands within the range of 3.33$-$3.21 $\mu$m in the experimental gas phase infrared spectra of some small PAH molecules with bay-type hydrogens in their molecular structures. 
The aromatic C$-$H stretching mode of PAH has slightly different frequencies depending on how many of the edge H atoms are in ``bay'' or ``non-bay'' configurations \citep{baus09} and it is possible to separate these components by imaging spectroscopic observations \citep{candian2012}.  However, the peak wavelengths of the ``bay'' and ``non-bay''modes are at 3.17--3.27 and 3.26--3.27 $\mu$m, respectively, both  shortward of the astronomical wavelengths of the 3.28 and 3.30 $\mu$m components derived by \citet{candian2012} by fitting the asymmetric profile of the 3.3 $\mu$m feature.

The aromatic C$-$H stretch at 3.3 $\mu$m and aliphatic C$-$H stretch at 3.4 $\mu$m is commonly seen in many carbonaceous compounds \citep{cataldo2020}.  There are also olefinic C$-$H stretching modes in this wavelength region.  
For example,  \citet{Chiar13} interpret the 3.28 $\mu$m component as the stretching mode of olefinic C$-$H bonds in amorphous hydrocarbons. The possibility that the 3.30 $\mu$m subfeature is due to olefinic C$-$H stretch is discussed by \citet{sadjadi2017}.

\subsection{The 6.2, 7.7, and 8.6 $\mu$m bands}
Based on a comparison between the astronomical UIE spectra with the laboratory spectrum of coronene, \citet{leger1984} assigns the 6.2 $\mu$m band to C--C stretch and the 8.6 $\mu$m to C--H in-plane bending mode.  However, the origin of the 7.7 $\mu$m feature is not clear \citep{allamandola1985}.   From the laboratory spectrum of coronene, the  7.4 and 6.4 $\mu$m features are suggested to be the combination of aromatic C$-$C stretch and C$-$H in plane bending  modes \citep{langhoff1996}.  
In later discussions under the PAH hypothesis, infrared bands between 6.1 and 6.5 $\mu$m are suggested to be due to pure aromatic C$-$C stretching modes, bands from 6.5 to 8.5 $\mu$m are due to coupled C--C stretching and C--H in-plane bending modes, and bands between 8.3 to 8.9 $\mu$m are due to C--H in-plane bending modes \citep{peeters2011}.
Since the peak wavelengths of PAH features shift with size of the molecule, the wavelength difference between 6.2 and 7.7 $\mu$m features have been used to infer the size of the PAH molecules \citep{hudgins1999, baus2008, baus09}.
 
However, all C--C stretching modes are very weak in comparison to the C--H modes in neutral PAH molecules. In order to explain the prominence of the features, PAH ions are suggested to be responsible for the UIE features in the 6$-$9 $\mu$m region \citep{hudgins2004} (Figure \ref{hudgins}). 
The need by the PAH hypothesis to simultaneously rely on neutral PAHs to explain the 3.3 and 11.3 $\mu$m features and PAH ions to explain the 6.2, 7.7, and 8.6 $\mu$m features poses difficulties in explaining the near universal presence of all these features in astronomical sources of very different UV backgrounds (Fig.~\ref{uchida}).

As the molecular structure gets more complex (e.g. in MAONs), many of the vibrational modes in the 6--10 $\mu$m region become coupled and cannot be identified as a single vibrational mode \citep{sadjadi2016}.

\subsection{The 11.3 $\mu$m band}

Soon after the discovery of the UIE bands, the 11.3 $\mu$m feature was identified as C--H out-of-plane bending mode of aromatic compounds \citep{duley1981}.
The observed peak position of the 11.3\,$\mu$m feature in astronomical sources is well defined  and  does not vary much in wavelength.  %Its profiles are also quite uniform in different objects, showing an asymmetric shape with a steep blue rise and a red tail \citep{vp04}.  
The 11.3 $\mu$m  feature  is also detected in absorption \citep{bh00},  and its wavelengths and profiles closely resemble those seen in emission. The 11.3 $\mu$m feature has a distinctive asymmetric profile, having a steep decline in the short wavelength side and a gradual extended wing in the long wavelength side \citep{van04}.  Such asymmetric profiles are difficult to explain by gas-phase molecular emissions.  In order to explain the observed profile of the 11.3 $\mu$m feature in the PAH hypothesis, PAH molecules of different mass are needed:  high-mass PAHs will produce the short-wavelength side and low-mass  PAHs the long-wavelength side \citep{candian2015}.  

Furthermore, the peak positions of the C$-$H out-of-plane bending modes of PAH molecules can be quite different due to mode coupling, molecular structure, and charge states of the molecules \citep{hv01}.  Because of the large wavelength variations of the out-of-plane bending modes of PAH molecules, it is difficult to assign or match the observed astronomical feature to specific PAH molecules. 
% incorporate  internal hydrogen in  carbonaceous microparticles  \citep{bk90}. 
Ring deformation vibrational mode of small carbonaceous molecules such as ethylene oxide ($c$-C$_2$H$_4$O) has also been proposed to explain the narrow 11.3\,$\mu$m feature \citep{bernstein2009}.

\citet{sadjadi2015} show that a mixture of pure PAH molecules, even including units of different sizes, geometry and charged states, is unable to fit the astronomical spectra.  In order to fit the astronomical profile of the 11.3 $\mu$m feature, oxygen-containing molecules are needed.

\subsection{Plateau features}

The 8 and 12 $\mu$m plateau features were first detected in proto-planetary nebulae by {\it IRAS} LRS and {\it KAO}  observations \citep{kwok1989, buss1990}.   
\citet{li2012} assign these two plateau features as ``wings of C--C and C--H bands'' of PAH molecular vibrations.  If this is the case, then similar ``wings'' should be seen around the 3.3 $\mu$m feature but it is not generally observed.  Alternatively, these plateau features have been identified as superpositions of in-plane and out-of-plane bending modes emitted by a mixture of aliphatic side groups attached to aromatic rings \citep{kwok2001}. 

The 15--20 $\mu$m plateau feature has been detected in young stellar objects, compact H{\small II} regions, and planetary nebulae, and is especially strong in some proto-planetary nebulae \citep{zhang2010}.  The wavelength region of this broad feature suggests that it could arise from C--C skeleton vibrational modes \citep{vankerckhoven2000}.

\section{Synthesis pathways}
The circumstellar envelopes of evolved stars (from AGB stars through proto-planetary nebulae to planetary nebulae) are ideal laboratories to test models of molecular synthesis.  The sequential formation of molecules from C$_2$, C$_3$, CN, to HCN, HC$_3$N, HC$_5$N, C$_2$H$_2$, to C$_6$H$_6$ suggests a bottom-up scenario of molecular synthesis. PAH molecules can be the products of molecular synthesis \citep{frenklach1989}.  The linking of islands of aromatic rings by aliphatic chains can lead to formation of MAONs.  
Fullerenes could be results of bottom-up processes 
\citep[e.g., from SiC grains,][]{bernal2019} or top-down process through the break up of MAONs  \citep{garcia2012, bernard2012, mic2012}.

%\section{Time scale of formation}
%The circumstellar environment can serve as a laboratory to study the synthesis of carriers of the UIE bands.  
The chemical time scales of molecular synthesis in the circumstellar environment is constrained  by the dynamical time scales of expanding circumstellar envelopes.  Gas-phase molecules and inorganic mineral solids are known to be forming in the circumstellar envelopes of AGB stars and planetary nebulae under low density and over very short time scales \citep{kwok2004}.
In proto-planetary nebulae, UIE bands are observed to emerge over time scales of $\sim$10$^3$ years  \citep{kwok1999}.  Infrared spectroscopic observations of novae following their outburst show that UIE bands emerge over time scales of weeks \citep{helton2011}.  In nova V705 Cas, the UIE bands, including a strong 3.4 $\mu$m feature, were seen within one year after outburst \citep{evans2005}.  These observations suggest that the abiotic synthesis of complex organics is extremely efficient in the circumstellar environment.

After their synthesis in circumstellar envelopes and their ejection into the interstellar medium, there is also a question of whether the carriers of UIE bands (whether they are PAH molecules or MAONs) can survive their journeys through the interstellar medium.  The aliphatic side groups in MAONs could be broken off by interstellar background radiation.  If multiple units of MAONs aggregate together they may resemble the macroscopic carbonaceous solids similar to the insoluble organic matter (IOM) observed in meteorites \citep{cody2011}.  Such solids are sturdy and are less likely to be destroyed by interstellar processes. Isotope anomalies in IOM suggest that it is probably of interstellar origin \citep{busemann2006}.

\section{Size and energetics}

One of the key questions about the carriers of the UIE bands is whether they are free-flying gas-phase molecules, nanoparticles, or solids.  Whereas the spectral properties of molecules (in particular PAH molecules) are relatively simple and well studied, the properties of nanoparticles consisting of $10^2-10^3$ heavy atoms cannot be easily extrapolated from bulk materials because of quantum surface effects.  
%Recent advances in computing have allowed quantum chemistry calculations for molecules with hundreds of carbon atoms  \citep{sadjadi2016}.  It is hoped that with such studies, one can learn more about the macro-molecular nanoparticles.

The astronomical UIE bands in planetary nebulae, H{\sc ii} regions, and galaxies are observed to lie on top of an infrared continuum of color temperature $\sim$10$^2$ K (Fig.~\ref{m82}).  In the diffuse interstellar medium, the color temperature of the infrared continuum is even lower.  Since the 3.3 $\mu$m band lies shortward (on the Wien's side) of the peak of the infrared continuum, it is unlikely that the 3.3 $\mu$m band is excited thermally.  Instead, stochastic heating by a single photon, exciting the carrier temporarily to a high temperature, is suggested to be the excitation mechanism  \citep{sellgren1984, omont1986}.

An alternate model of chemical heating was proposed by \citet{duley2011}, who suggest that this process can heat carbonaceous solid particles of sizes from 5$-$100 nm to emit the 3.3 $\mu$m band.  The release of H$_2$ molecules from chemical heating is also consistent with the correlation between observed 2.18 $\mu$m H$_2$ emission with the 3.3 $\mu$m band. 

%``Within the astrophysical context, the PAH family includes PAH related species such as, for example, PAHs with side groups, heterosubstituted PAHs, fully or partially (de)hydrogenated PAHs, and PAH clusters.'' \citep{peeters2021}.

If stochastic heating is indeed the excitation mechanism of UIE bands, then the sizes of the carrier must be limited to molecules or nanoparticles.  If chemical heating can work, then the size limit can be extended upwards.  In either case, we still have to work out the relationship between the UIE carriers and the micron-size grains that contribute to the underlying continuum (section \ref{continuum}).

\section{The need for a correct identification of the origin of the UIE bands}

Since the UIE bands are commonly used as a tracer of galactic evolution, it is pertinent that we correctly identify the chemical nature of the carrier of the bands. The UIE bands are detected in galaxies with redshifts $>$4 \citep{riechers2014, armus2020} and are particularly prominent in ultraluminuous infrared galaxies.  The power emitted in the UIE bands can be as high as 20\% of the total energy output of these galaxies \citep{smith2007}.  Given the nearly invariance of the band positions, the UIE bands have been used as redshift indicators \citep{elbaz2005}.  
In nearby galaxies, the distribution of the UIE bands can be mapped.  The  {\it AKARI} satellite has mapped the distribution of the 3.3 and 3.4 $\mu$m bands, showing emission from aromatic and aliphatic species are present in the halo of M82, as far as 2 kpc from the galactic center \citep{yamagishi2012}.

Under the assumption that UIE bands are PAH molecules excited by UV photons, the UIE bands have been used as tracers of star formation in galaxies \citep[see e.g.,][]{smith2007,gt08,wh10, li2020} as well as the star formation activities as a fraction of the total infrared luminosities of galaxies  \citep{peeters2004, spoon2004}.  Under similar assumptions, the UIE bands have also been used as tracers of elemental and chemical evolution of galaxies \citep{genzel1998}.

Based on the assumption that the 11.3 $\mu$m feature originates from neutral and large PAHs, and the  6.2, 7.7, and 8.6\,$\mu$m features originate  from PAH ions, the strength ratios between the 11.3\,$\mu$m feature and 6 to 9 $\mu$m UIE features  have been used to estimate the properties of the background radiation fields \citep{gt08, berne2009}. These results are used to determine the star formation rates of galaxies \citep{calzetti}.
The physical validity of these studies, however, depends on a correct interpretation of the origin of the UIE bands.

\section{Conclusions}

The family of astronomical UIE bands is a rich spectral phenomenon which carrier is yet to be unambiguously identified.  It is likely to be a carbonaceous substance whose structure is more complicated than it is often assumed.
Although the PAH hypothesis has been popular in the astronomical community, its definition has been evolving from the original chemical definition of planar, pure carbon and hydrogen ring molecules, to a collection of molecules including 
%``Within the astrophysical context, the PAH family includes PAH related species such as, for example, 
``PAHs with side groups, heterosubstituted PAHs, fully or partially (de)hydrogenated PAHs, and PAH clusters.'' \citep{peeters2021}.  This revision of definition is moving the PAH hypothesis closer to other models of the UIE bands.  

The debate between the PAH and HAC/MAON models is more than a question of semantics as there are fundamental differences between these two models:  Do the UIE carriers contain pure rings or mixed rings and chains?  Are their geometry 2-D or 3-D?  Are they pure C$-$H compounds or contain impurities?  Are their structure regular or amorphous?  Are the carriers gas-phase free-flying molecules or solids?  If answers to these questions are toward the latter, we have to call these carriers by their proper chemical terminology.

Since the UIE phenomenon is seen throughout the Universe, even during its early epochs, a correct identification of the carrier is of great importance.  Due to the strengths of the UIE bands, the carrier represents a major reservoir of carbon. If the carrier of UIE bands are PAH molecules, they would contain $\sim$10\% of cosmic carbon \citep{peeters2021}.  Many current models of the interstellar medium are based the premise that PAH molecules are the dominant factors in  photoelectric heating of interstellar gas and  in the ionization balance inside molecular clouds.  Whether the carriers of the UIE bands are a collection of free-flying gas-phase PAH molecules or complex organic solid particles has significantly different implications on our understanding of cosmic chemical synthesis, energy exchange between stars and the interstellar medium,  and galactic chemical enrichment.  Further computational and experimental studies of the vibrational properties of amorphous carbonaceous compounds  are needed to solve the UIE mystery.

\acknowledgments
I thank SeyedAbdolreza Sadjadi for producing Figure 7.   I also thank Larry Bernstein, Franco Cataldo, Anibal Garc\'{i}a-Hern\'{a}ndez, Louis d'Hendecourt,  Walt Duley, Roger Knacke, Takashi Onaka, Ray Russell, Alan Tokunaga, and Yong Zhang for helpful discussions,  as well as the two anonymous referees who provided helpful comments leading to an improvement the paper.
This work is supported by a grant from the Natural and Engineering Research Council of Canada.


\begin{thebibliography}{}
\bibitem[Allamandola et al.(1985)]{allamandola1985} 
Allamandola, L. J., Tielens, A. G. G. M., \& Barker, J. R.:  \apj\ {\bf 290}, L25 (1985)

\bibitem[Allamandola et al.(1989)]{all89}  
Allamandola, L. J., Tielens, A. G. G. M., and Barker, J. R.:  1989, \apjs\ {\bf 71}, 733 (1989)

\bibitem[Allamandola et al.(1999)]{all99}  
Allamandola, L. J., Hudgins, D. M., and Sandford, S. A.:  \apjl\ {\bf 511}, L115 (1999)



\bibitem[Armus et al.(2020)]{armus2020}
Armus, L., Charmandaris, V., \& Soifer, B. T.: Nature Astronomy, {\bf 4}, 467 (2020)

\bibitem[Barker et al.(1987)]{barker} 
Barker, J.~R., Allamandola, L.~J.,  Tielens, A.~G.~G.~M.: \apjl\ {\bf 315}, L61 (1987)

\bibitem[Bauschlicher et al.(2008)]{baus2008}
Bauschlicher, C. W., Jr., Peeters, E., Allamandola, L. J.:  \apj\ {\bf 678}, 316 (2008)

\bibitem[Bauschlicher et al,(2009)]{baus09} 
Bauschlicher, C. W., Peeters, E.,  Allamandola, L. J.:  \apj\ {\bf 697}, 311 (2009)

\bibitem[Bauschlicher et al.(2010)]{bb10} 
Bauschlicher, C.~W., Jr., Boersma, C., Ricca, A., et al.:  \apjs\ {\bf 189}, 341 (2010)

%\bibitem[Bern{\'e} et al.(2007)]{ber07} Bern{\'e}, O., Joblin, C., Deville, Y., et al.\ 2007, \aap, 469, 575 

\bibitem[Bernal et al.(2019)]{bernal2019}
Bernal, J. J., Haenecour, P., Howe, J., et al.: \apj, {\bf 883}, L43 (2019)

\bibitem[Bern\'{e} et al.(2009)]{berne2009} 
Bern\'{e}, O., Fuente, A., Goicoechea, J. R., et al.: \apj\ {\bf 706}, L160 (2009)

%\bibitem[Boersma et al.(2013)]{bb13} 
%Boersma, C., Bregman, J.~D.,  Allamandola, L.~J.: \apj\ {\bf 769}, 117 (2013)

\bibitem[Boersma et al.(2014)]{bb14} 
Boersma, C., Bauschlicher, C.~W., Jr., Ricca, A., et al.: \apjs\ {\bf 211}, 8 (2014) 


%\bibitem[Bernard-Salas et al.(2009)]{bernard-salas2009} 
%Bernard-Salas, J., Peeters, E., Sloan, G. C., et al.: \apj\ {\bf 699}, 1541 (2009)

\bibitem[Bernard-Salas et al.(2012)]{bernard2012}
Bernard-Salas, J., Cami, J., Peeters, E., Jones, A.P., Micelotta, E.R., Groenewegen, M.A.T. : \apj\ {\bf 757}, 41 (2012)

\bibitem[Bernstein and Lynch(2009)]{bernstein2009}
Bernstein, L.S., Lynch, D.K.:  Astrophys J {\bf 704}, 226 (2009) 

%\bibitem[Bradley et al.(2005)]{bradley2005}
%Bradley, J., Dai, Z.R., Erni, R., Browning, N., Graham, G., Weber, P., Smith, J., Hutcheon, I., Ishii, H., Bajt, S., Floss, C., Stadermann, F., Sandford, S.: Science   {\bf 307}. 244 (2005)

\bibitem[Bregman et al.(2000)]{bh00} 
Bregman, J.~D., Hayward, T.~L., Sloan, G.~C.: \apjl\ {\bf 544}, L75  (2000)

\bibitem[Burkhardt et al.(2021)]{burkhardt2021}
Burkhardt, A. M., Lee, K.L.K., Changala, B., et al.: \apj, {\bf 913}, L18 (2021)

\bibitem[Buss et al.(1990)]{buss1990} 
Buss, R. H., Cohen, M., Tielens, A., et al.: \apj\ {\bf 365}, L23 (1990)

\bibitem[Busemann et al.(2006)]{busemann2006}
Busemann, H., Young, A. F., Alexander, C. M. O. D., et al.: Science, {\bf 312}, 727 (2006)

\bibitem[Calzetti(2011)]{calzetti}
Calzetti, D.: In: Joblin, C. Tielens, A.G.G.M. (eds) PAHs and the Universe,   EAS Publications Series {\bf 46}, p.~133 (2011)


\bibitem[Cami(2011)]{cami2011}
Cami, J.: In: Joblin, C. Tielens, A.G.G.M. (eds) PAHs and the Universe,   EAS Publications Series {\bf 46}, p.~117 (2011)

\bibitem[Cami et al.(2010)]{cami2010}
Cami, J., Bernard-Salas, J., Peeters, E., Malek, S.E.: Science, {\bf 329},1180 (2010)

%\bibitem[Campbell et al.(2015)]{campbell2015}
%Campbell, E.K., Holz, M., Gerlich, D., Maier, J.P.:  Nature {\bf 523}. 322  (2015)

\bibitem[Candian and Sarre(2015)]{candian2015}
Candian, A., Sarre, P.J.:  \mnras\ {\bf 448}, 2960 (2015)

\bibitem[Candian et al.(2012)]{candian2012} 
Candian, A., Kerr, T. H., Song, I.-O., McCombie, J., Sarre, P. J.: \mnras\ {\bf 426}, 389 (2012)

\bibitem[Carpentier et al.(2012)]{car2012}
Carpentier, Y., et al.: \aap\ {\bf 548}, 40 (2012)

\bibitem[Cataldo(2004)]{cataldo2004}
Cataldo, F:  In: P. Ehrenfreund et al. (eds) Astrobiology: Future Perspective, p.~97  Kluwer (2004)


\bibitem[Cataldo et al. (2020)]{cataldo2020}
Cataldo, F., Garc\'{i}a-Hern\'{a}ndez, D. A.,  Manchado, A.: \apss\ {\bf 365}, 81 (2020)

%\bibitem[Candian et al.(2012)]{candian2012}
%Candian, A., Kerr, T. H., Song, I.-O., McCombie, J., \& Sarre, P. J. 2012, \mnras, 426, 389



\bibitem[Cataldo et al.(2002)]{cataldo2002}
Cataldo, F., Keheyan, Y., Heymann, D.:   International J. of Astrobiology {\bf 1}, 79 (2002)

%\bibitem[Cataldo, Garc\'{i}a-Hern\'{a}ndez and Manchado(2020)]{cataldo2020}
%Cataldo, F., Garc\'{i}a-Hern\'{a}ndez, D. A., \& Manchado, A.: \apss  {\bf 365}. 81 (2020)

\bibitem[Cernicharo et al.(2001)]{cernicharo2001}
Cernicharo, J., Heras, A. M., Tielens, A. G. G. M., et al.:  \apj\ {\bf 546}, L123 (2001)

\bibitem[Cernicharo et al.(2021)]{cernicharo2021}
Cernicharo, J., Ag\'{u}ndez, M., Cabezas, C., et al.: \aap, {\bf 649}, L15 (2021)

%\bibitem[Chang et al.(2006)]{chang2006}
%Chang, H.-C., Chen, K., Kwok, S.:  Astrophys. J. Lett. {\bf 639}. L63 (2006)

\bibitem[Chan et al.(2001)]{chan2001}
Chan, K.-W., Roellig, T. L., Onaka, T., et al.: \apj, {\bf 546}, 273 (2001)

\bibitem[Chiar et al.(2000)]{chiar2000}
Chiar, J. E., Tielens, A. G. G. M., Whittet, D. C. B., et al.: \apj\ {\bf 537}, 749 (2000)

\bibitem[Chiar et al.(2013)]{Chiar13}
Chiar, J. E., Tielens, A. G. G. M., Adamson, A. J., Ricca, A.: \apj\ {\bf 770}, 78 (2013)



\bibitem[Chung and Violi(2011)]{chung2011}
Chung, S-H., Violi, A.: Proc. Combust. Inst., {\bf 33}, 693 (2011)

\bibitem[Ciar(1964)]{ciar1964}
Ciar, E.: Polycyclic Hydrocarbons (New York: Academic Press) (1964)

\bibitem[Clayton et al.(2003)]{clayton2003} 
Clayton, G. C., Gordon, K. D., Salama, F., et al.:  \apj\ {\bf 592}, 947 (2003)

%\bibitem[Cohen et al.(1975)]{cohen1975} 
%Cohen, M., Anderson, C. M., Cowley, A., et al.: \apj\ {\bf 196}. 179 (1975)

\bibitem[Cody et al.(2011)]{cody2011} 
Cody, G. D., Heying, E., Alexander, C.M.O., Nittler, L.R., Kilcoyne, A.L.D., Sandford, S.A., Stroud, R.M.: Proceedings of the National Academy of Sciences of the U.S.A. {\bf 108}, 19171 (2011)


\bibitem[Cohen et al.(1986)]{cohen1986}
Cohen, M., Allamandola, L., Tielens, A. G. G. M., et al.:  \apj\ {\bf 302}, 737 (1986)

\bibitem[Cohen et al.(1989)]{cohen1989}
Cohen, M., Tielens, A. G. G. M., Bregman, J., et al.: \apj\ {\bf 341}, 246 (1989)

\bibitem[Colangeli et al.(1997)]{col1997}
Colangeli, L., Bussoletti, E., Pestellini, C. C., Mennella, V., Palomba, E., Palumbo, P., Rotundi, A.:  Advances in Space Research, {\bf 20}, 1617 (1997)

\bibitem[Cook and Saykally(1998)]{cook1998}
Cook, D. J., Saykally, R. J.:  \apj\ {\bf 493}, 793 (1998)

\bibitem[Cook et al.(1996)]{cook1996}
Cook, D. J. et al.:  \nat\ {\bf 380}, 227 (1996)



\bibitem[Dartois et al.(2004a)]{dartois2004a}
Dartois, E., Marco, O., Mu\~{n}oz-Caro, G. M., et al.: \aap\ {\bf 423}, 549 (2004a)

\bibitem[Dartois(2004b)]{dartois2004b}
Dartois, E., Mu\~{n}oz Caro, G. M., Deboffle, D.,  d'Hendecourt, L.:  \aap\ {\bf 423}, L33 (2004b)

\bibitem[Dartois(2011)]{dartois2011}
Dartois, E.: In: Joblin, C. Tielens, A.G.G.M. (eds) PAHs and the Universe, EAS Publications Series {\bf 46}, p.~381  (2011)

\bibitem[Dischler et al.(1983)]{dischler1983}
Dischler, B., Bubenzer, A.,  Koidl, P.: Appl.~Phys.~Lett. {\bf 42}, 636 (1983)

\bibitem[Donn(1968)]{donn1968}
Donn, B.: \apjl\ {\bf 152}, L129 (1968)

\bibitem[Donn et al.(1989)]{donn1989}
Donn, B., Allen, J., Khanna, R.:  In: Allamandola, L.J., Tielens, A.G.G.M. (eds) Interstellar Dust,  p.~181 (1989)

\bibitem[Draine \& Li(2007)]{dra07}  
Draine, B. T., and Li, A.: \apj\ {\bf 657}, 810 (2007)


\bibitem[Draine et al.(2021)]{draine2021}
Draine, B. T., Li, A., Hensley, B. S., et al.: \apj, {\bf 917}, 3 (2021)

\bibitem[Duley(1993)]{duley1993}
Duley, W. W.: In: Kwok, S. (ed) Astronomical Infrared Spectroscopy: Future Observational Directions. ASP Conf. Ser. vol. 41,  p.~241 (1993)

\bibitem[Duley(2000)]{duley2000} 
Duley, W. W.: \apj\ {\bf 528}, 841 (2000)

\bibitem[Duley and Williams(1979)]{duley1979} 
Duley, W. W., Williams, D. A.: \nat\ {\bf 277}, 40 (1979)

\bibitem[Duley and Williams(1981)]{duley1981}
Duley, W.W., Williams, D.A.: Mon. Not. R. Astron. Soc. {\bf 196}, 269 (1981)

\bibitem[Duley and Williams(2011)]{duley2011}
Duley, W. W.,  Williams, D. A.: \apjl\ {\bf 737}, L44 (2011)

\bibitem[Duley et al.(2005)]{duley2005}
Duley, W. W., Lazarev, S., Scott, A.: \apjl\ {\bf 620}, L135 (2005)

\bibitem[Elbax et al.(2005)]{elbaz2005}
Elbaz, D., Le Floc'h, E., Dole, H.,  Marcillac, D.:  \aap\ {\bf 434}, L1 (2005)

\bibitem[Endo et al.(2021)]{endo2021}
Endo, I., Sakon,I.,  Onaka, T., Kimura,Y.,  Kimura,S.,  Wada, S. Helton, L.A., Lau, R.M.,  Kebukawa, Y.,  Muramatsu, Y.,  Ogawa, N.O., Ohkouchi,N., Nakamura, M., Kwok, S.: \apj, {\bf 917}, 103 (2021)

%\bibitem[El\'{i}asd\'{o}ttir et al.(2009)]{elias2009}
%El\'{i}asd\'{o}ttir, \'{A}. et al.: Astrophys. J. {\bf 697}, 1725 (2009)

\bibitem[Evans et al.(2005)]{evans2005} 
Evans, A., Tyne, V. H., Smith, O., et al.: \mnras\ {\bf 360}, 1483 (2005)

%\bibitem[Foing and Ehrenfreund(1994)]{foing1994}
%Foing, B.H., Ehrenfreund, P.: Nature {\bf 369}. 296 (1994)

\bibitem[Frenklach \& Feigelson(1989)]{frenklach1989}
Frenklach, M., \& Feigelson, E. D.: \apj, {\bf 341}, 372 (1989)

\bibitem[Gadallah et al.(2012)]{gad2012}
Gadallah, K. A. K., Mutschke, H.,  J\"{a}ger, C.: \aap\ {\bf 544}, 107 (2012)

\bibitem[Galliano et al.(2008)]{gt08} 
Galliano, F., Madden, S.~C., Tielens, A.~G.~G.~M., Peeters, E., Jones, A.~P.: \apj\ {\bf 679}, 310 (2008)

\bibitem[Garc\'{i}a-Hern\'{a}ndez et al.(2010)]{garcia2010} 
Garc\'{i}a-Hern\'{a}ndez, D.A., Manchado, A., Garc\'{i}a-Lario, P., et al.: \apj\ {\bf 724}, L39 (2010)

\bibitem[Garc\'{i}a-Hern\'{a}ndez et al.(2012)]{garcia2012}
Garc\'{i}a-Hern\'{a}ndez, D.A., Villaver, E., Garc\'{i}a-Lario, P., Acosta-Pulido, J.A., Manchado, A., Stanghellini, L., Shaw, R.A., Cataldo, F.: \apj {\bf 760}, 107 (2012)

\bibitem[Geballe et al.(1992)]{geballe1992} 
Geballe, T. R., Tielens, A. G. G. M., Kwok, S., Hrivnak, B. J.:  \apjl\ {\bf 387}, L89 (1992)

\bibitem[Genzel et al.(1998)]{genzel1998}
Genzel, R., Lutz, D., Sturm, E., et al.: \apj\ {\bf 498}, 579 (1998)

\bibitem[Gillett et al.(1967)]{gillett1967}
Gillett, F. C., Low, F. J., Stein, W. A.: \apjl\ {\bf 149}, L97 (1967)

\bibitem[Gillett et al.(1973)]{gillett1973} 
Gillett, F. C., Forrest, W. J., Merrill, K. M.:  \apj\ {\bf 183}, 87 (1973)

\bibitem[Godard et al.(2011)]{god2011}
Godard, M., et al.: \aap\ {\bf 529}, 146 (2011)

\bibitem[Goto et al.(2003)]{Goto2003}
Goto, M., Gaessler, W., Hayano, Y., et al.: \apj\ {\bf 589}, 419 (2003)

\bibitem[Goto et al.(2007)]{goto2007} 
Goto, M., Kwok, S., Takami, H., et al.: \apj\ {\bf 662}, 389 (2007)

\bibitem[Gredel et al.(2011)]{gredel2011} 
Gredel, R., Carpentier, Y., Rouill\'{e}, G., et al.: \aap\ {\bf 530}, 26 (2011)

\bibitem[Guillois et al.(1996)]{guillois1996}
Guillois, O., Nenner, I., Papoular, R., Reynaud, C.: \apj\ {\bf 464}, 810 (1996)


\bibitem[Helton et al.(2011)]{helton2011}
Helton, L. A., Evans, A., Woodward, C. E.,  Gehrz, R. D.: In: Joblin, C. Tielens, A.G.G.M. (eds) PAHs and the Universe,    EAS Publications Series 46, p.~407 (2011)

\bibitem[Herlin et al.(1998)]{herlin1998} 
Herlin, N., Bohn, I., Reynaud, C., Cauchetier, M., Galvez, A.,  Rouzaud, J.-N.:  \aap\ {\bf 330}, 1127 (1998)

%\bibitem[Hsia et al.(2016)]{hsia2016}
%Hsia, C.-H., Sadjadi, S., Zhang, Y., \& Kwok, S.: \apj\ {\bf 832}, 213 (2016)

\bibitem[Holmlid(2000)]{holmlid2000}
Holmlid, L.: \aap\ {\bf 358}, 276 (2000)

\bibitem[Hony et al.(2001)]{hv01} 
Hony, S., Van Kerckhoven, C., Peeters, E., et al.: \aap\ {\bf 370}, 1030 (2001)


\bibitem[Hoyle and Wickramasinghe(1977)]{hoyle1977}
Hoyle, F., Wickramasinghe, N.C.: Nature, {\bf 268} 610 (1977)


%\bibitem[Hrivnak et al.(2000)]{hrivnak2000} 
%Hrivnak, B. J., Volk, K., Kwok, S.: \apj\ {\bf 535}, 275 (2000)

\bibitem[Hrivnak et al.(2007)]{hrivnak2007} 
Hrivnak, B. J., Geballe, T. R., Kwok, S.:  \apj\ {\bf 662}, 1059 (2007)



\bibitem[Hu and Duley(2008)]{hu2008}
Hu, A., Duley, W.W.: Astrophys. J. Lett. {\bf 677},  L153 (2008)

\bibitem[Hu et al.(2006)]{hu2006} 
Hu, A., Alkhesho, I.,  Duley, W. W.: \apj\ {\bf 653}, L157 (2006)

\bibitem[Hudgins and Allamandola(2004)]{hudgins2004}
Hudgins, D. and Allamandola, L.J.:
%Polycyclic aromatic hydrocarbons and infrared astrophysics: The state of the PAH model and a possible tracer of nitrogen in carbon-rich dust. 
In: Witt, A.N., Clayton, G.C., and Draine, B.T. (eds.) Astrophysics of Dust,   ASP Conf. Ser. vol. 309, p.~ 665 (2004)

\bibitem[Hudgins and Allamandola(1999)]{hudgins1999}
Hudgins, D.M., and Allamandola, L. J.: \apjl\  {\bf 513} L69 (1999)

\bibitem[Imanishi(2000)]{imanish2000}
Imanishi, M.: \mnras\ {\bf 319}, 331 (2000)

\bibitem[Imanishi et al.(2010)]{imanishi2010}
Imanishi, M., Nakagawa, T., Shirahata, M., Ohyama, Y., Onaka, T.:  Astrophys. J. {\bf 721}, 1233 (2010)

\bibitem[J\"{a}ger et al.(2008)]{jager2008} 
J\"{a}ger, C., Mutschke, H., Henning, T., \& Huisken, F.: \apj\ {\bf 689}, 249 (2008)

%\bibitem[Joblin \& Tielens(2011)]{pah}
%Joblin, C., \& Tielens, A.G.G.M.: PAHs and the Universe, EAS Publications Series 46 (2011)

\bibitem[Joblin et al.(1995)]{joblin1995}
Joblin, C., Boissel, P., Leger, A., D'Hendecourt, L., Defourneau, D.:  \aap\ {\bf 299}, 835 (1995)

\bibitem[Jones(2012a)]{jones2012a}
Jones, A. P.: \aap\ {\bf 540}, 1 (2012a)

\bibitem[Jones(2012b)]{jones2012b}
Jones, A. P.: \aap\ {\bf 540}, 2 (2012b)
%\bibitem[Jones(2012c)]{jones2012c}
%Jones, A. P. 2012c, \aap, 545, 3

%\bibitem[Jones (2016)]{jones2016}
%Jones, A. P.: Royal Society Open Science, 3 (2016)

\bibitem[Jones et al.(1990)]{jones1990} 
Jones, A. P., Duley, W. W.,  Williams, D. A.:  \qjras\ {\bf 31}, 567 (1990)

\bibitem[Jones et al.(2013)]{jones2013}
Jones, A. P., Fanciullo, L., K\"{o}hler, M., et al.: \aap\ {\bf 558}, 62 (2013)



\bibitem[Jourdain de Muizon et al.(1986)]{Jourdain86} 
Jourdain de Muizon, M., Geballe, T. R., D'Hendecourt, L. B., Baas, F.: \apjl\ {\bf 306}, L105 (1986)

\bibitem[Jourdain de Muizon et al.(1990)]{deMuizon1990} 
Jourdain de Muizon, M., D'Hendecourt, L. B.,  Geballe, T. R.:  \aap\ {\bf235}, 367 (1990)

\bibitem[Kahanp\"{a}\"{a} et al.(2003)]{kahanpaa2003} 
Kahanp\"{a}\"{a}, J., Mattila, K., Lehtinen, K., et al.: \aap\ {\bf 405}, 999 (2003)

\bibitem[Khare and Sagan(1973)]{khare1973}
Khare, B. N., Sagan, C.:  \icarus\ {\bf 20}, 311 (1973)

\bibitem[Knacke(1977)]{knacke1977}
Knacke, R.F.:  Nature {\bf 269}, 132 (1977)

\bibitem[Kwok(2004)]{kwok2004} 
Kwok, S.:  \nat\ {\bf 430}, 985 (2004)

\bibitem[Kwok(2007)]{kwok2007}
Kwok, S.: Physics and Chemistry of the Interstellar Medium, University Science Books, Sausalito (2007)

\bibitem[Kwok(2016)]{kwok2016}
Kwok, S.: \aapr\ {\bf 24}, 8 (2016)

\bibitem[Kwok and Zhang(2011)]{kz2011}
Kwok, S., Zhang, Y.: Nature, {\bf 479}, 80 (2011)

\bibitem[Kwok and Zhang(2013)]{kz2013}
Kwok, S., Zhang, Y.: Astrophys. J. {\bf 771},  5 (2013)

\bibitem[Kwok et al.(2001)]{kwok2001}
Kwok, S., Volk, K., Bernath, P.: Astrophys. J. Lett. {\bf 554},  L87 (2001)


\bibitem[Kwok et al.(1989)]{kwok1989} 
Kwok, S., Volk, K. M., \& Hrivnak, B. J.: Astrophys. J. Lett. {\bf 345}, L51 (1989)


\bibitem[Kwok et al.(1999)]{kwok1999} 
Kwok, S., Volk, K.,  Hrivnak, B. J.: \aap\ {\bf 350}, L35 (1999)

\bibitem[Langhoff(1996)]{langhoff1996}
Langhoff, S. R.: The Journal of Physical Chemistry  {\bf 100}, 2819 (1996)

\bibitem[L\'{e}ger and Puget(1984)]{leger1984}
L\'{e}ger, A., Puget, J.L.: Astron. Astrophys. {\bf 137}, L5 (1984)

\bibitem[L\'{e}ger et al.(1989)]{leger1989}
L\'{e}ger, A., Verstraete, L., D'Hendecourt, L., et al.:
%The PAH Hypothesis and the Extinction Curve, in Proceedings of the 135th Symposium of the International Astronomical Union, held in Santa Clara, California, 26-30 July 1988. Edited by Louis J. Allamandola and A. G. G. M. Tielens. 
In: Allamandola, L.J., and Tielens, A.G.G.M. (eds) IAU Symposium 135 Interstellar Dust.  Kluwer Academic Publishers, Dordrecht, p.173 (1989)

\bibitem[Li(2020)]{li2020}
Li, A.: Nature Astronomy, {\bf 4}, 339 (2020)

\bibitem[Li \& Draine(2001)]{li2001}
Li, A., \& Draine, B. T.: \apj\ {\bf 554}, 778 (2001)


\bibitem[Li and Draine(2012)]{li2012}
Li, A., and Draine, B. T. \apjl\ {\bf 760}, L35 (2012)


\bibitem[Mattioda et al.(2020)]{mattioda2020}
Mattioda, A. L., Hudgins, D. M., Boersma, C., et al.: \apjs\ {\bf 251}, 22 (2020)

\bibitem[McGuire et al.(2021)]{mcguire2021}
McGuire, B. A., Loomis, R. A., Burkhardt, A. M., et al.: Science {\bf 371}, 1265 (2021)

%\bibitem[Mennella et al.(1996)]{men1996}
%Mennella, V., Colangeli, L., Palumbo, P., Rotundi, A., Schutte, W., \& Bussoletti, E.: \apjl\ {\bf 464}, L191 (1996)

\bibitem[Mennella et al.(1999)]{mennella1999} 
Mennella, V., Brucato, J. R., Colangeli, L., \& Palumbo, P.: \apjl\ {\bf 524}, L71 (1999)

\bibitem[Mennella et al.(2003)]{mennella2003}
Mennella, V., Baratta, G. A., Esposito, A., Ferini, G., \& Pendleton, Y. J.: \apj\ {\bf 587}, 727 (2003)


%\bibitem[Mennella et al.(1998)]{Mennella1998}
%Mennella, V., Colangeli, L., Bussoletti, E., Palumbo, P., Rotundi, A.: Astrophys. J. Lett. {\bf 507}. L177 (1998)

\bibitem[Merrill et al.(1975)]{merrill1975} 
Merrill, K. M., Soifer, B. T., Russell, R. W.:  \apjl\ {\bf 200}, L37 (1975)

\bibitem[Micelotta et al.(2012)]{mic2012}
Micelotta, E., Jones, A.P., Peeters, E., Bernard-Salas, J., \& Fanchini, G.: \apj\ {\bf 761}, 35 (2012)


\bibitem[Omont(1986)]{omont1986}
Omont, A.: \aap\ {\bf 164}, 159 (1986)

\bibitem[Otsuka et al.(2013)]{otsuka2013}
Otsuka, M., Kemper, F., Hyung, S., Sargent, B.A., Meixner, M., Tajitsu, A., Yanagisawa, K.:
\apj\, {\bf 764}, 77 (2013)


\bibitem[Papoular et al.(1989)]{papoular1989}
Papoular, R., Conrad, J., Giuliano, M., Kister, J., Mille, G.: 
%(1989) A coal model for the carriers of the unidentified IR bands. 
Astron. Astrophys. {\bf 217},  204 (1989)

\bibitem[Papoular(2001)]{papoular2001}
Papoular, R.: Astron. Astrophys. {\bf 378}, 597 (2001)



\bibitem[Peeters(2011)]{peeters2011}
Peeters, E.: In: Cernicharo, J. \& Bachiller, R. (eds) IAU Symposium 280 The Molecular Universe. CUP,  p.~149 (2011)


\bibitem[Peeters et al.(2002)]{peeters2002}
Peeters, E. et al.:  \aap\  {\bf 390}, 1089 (2002)

\bibitem [Peeters et al.(2004)]{peeters2004}
Peeters, E., Spoon, H. W. W.,  Tielens, A. G. G. M.:  \apj\ {\bf 613}, 986 (2004)

\bibitem[Peeters et al.(2012)]{peeters2012}
Peeters, E., Tielens, A. G. G. M., Allamandola, L. J., Wolfire, M. G.: 2012, \apj\ {\bf 747}, 44 (2012)

\bibitem[Peeters et al.(2021)]{peeters2021}
Peeters, E., Mackie, C., Candian, A., Tielens, A. G. G. M.: Acc.~Chem.~Res. {\bf 54}, 1921 (2021)



\bibitem[Pendleton et al.(1994)]{pendleton1994}
Pendleton, Y. J., Sandford, S. A., Allamandola, L. J., et al.:  \apj\ {\bf 437}, 683 (1994)

\bibitem[Puetter et al.(1979)]{puetter1979}
Puetter, R.C., Russell, R.W., Willner, S.P., Soifer, B.T.: %(1979) Spectrophotometry of compact H II regions from 4 to 8 microns. 
Astrophys. J. {\bf 228}, 118 (1979)

\bibitem[Riechers et al.(2014)]{riechers2014}
Riechers, D. A., Pope, A., Daddi, E., et al.:  \apj\ {\bf 786}, 31 (2014)


\bibitem[Robertson(2002)]{robertson2002}
Robertson, J.: 
%(2002) Diamond-like amorphous carbon. 
Materials Science and Engineering R{\bf 37}, 129 (2002)


\bibitem[Russell et al.(1977a)]{russell1977a}
Russell, R. W., Soifer, B. T., \& Merrill, K. M.:  \apj\ {\bf 213}, 66	(1977a)

\bibitem[Russell et al.(1977b)]{russell1977b}
Russell, R.W., Soifer, B.T., Willner, S.P.:
% (1977) The 4 to 8 micron spectrum of NGC 7027. 
Astrophys. J. Lett. {\bf 217}, L149 (1977b)



\bibitem[Russell et al.(1978)]{russell1978} 
Russell, R. W., Soifer, B. T.,  Willner, S. P.  \apj\ {\bf 220}, 568 (1978)

\bibitem[Sagan and Khare(1979)]{sagan1979} 
Sagan, C., Khare, B. N.: \nat\ {\bf 277}, 102 (1979)


\bibitem[Sadjadi et al.(2015)]{sadjadi2015}
Sadjadi, S., Zhang, Y.,  Kwok, S.:  \apj\ {\bf 807}, 95 (2015)

\bibitem[Sadjadi et al.(2016)]{sadjadi2016}
Sadjadi, S., Kwok, S.,  Zhang, Y.: IOP Journal of Physics: conference series {\bf 728}, 062003 (2016)

\bibitem[Sadjadi et al.(2017)]{sadjadi2017}
Sadjadi, S., Zhang, Y., Kwok, S.: \apj\ {\bf 845}, 123 (2017)

\bibitem[Sakata et al.(1983)]{sakata1983}
Sakata, A., Wada, S., Okutsu, Y., et al.: \nat\ {\bf 301}, 493 (1983)


\bibitem[Sakata et al.(1984)]{sakata1984} 
Sakata, A., Wada, S., Tanabe, T.,  Onaka, T. 1984, \apjl\ {\bf 287}, L51 (1984)

\bibitem[Sakata et al.(1987)]{sakata1987}
Sakata, A., Wada, S., Onaka, T., Tokunaga, A.T.:
%Infrared spectrum of quenched carbonaceous composite (QCC). II - A new identification of the 7.7 and 8.6 micron unidentified infrared emission bands. 
Astrophys. J. Lett. {\bf 320}. L63 (1987)



\bibitem[Sakata et al.(1990)]{sakata1990}
Sakata, A., Wada, S., Onaka, T., \& Tokunaga, A. T.: \apj\ {\bf 353}, 543 (1990)

%\bibitem[Sakata etal(1992)]{sakata1992}
%Sakata, A. etal: Astrophys. J. {\bf 393}. L83 (1992)

\bibitem[Salama(2008)]{salama2008}
Salama, F.: In: Kwok, S. \& Sandford, S. (eds) IAU Symp 251 Organic Matter in Space, CUP, p.~357 (2008)


\bibitem[Salama et al.(2011)]{salama2011}
Salama, F., Galazutdinov, G. A., Krelowski, J., et al.: \apj\ {\bf 728}, 154 (2011)

\bibitem[Sandford et al.(1991)]{sandford1991} 
Sandford, S. A., Allamandola, L. J., Tielens, A. G. G. M., et al.: \apj\  {\bf 371}, 607 (1991)


%\bibitem[Sarre(2006)]{sarre2006}
%Sarre, P.J.:  J. Molecular Spectroscopy {\bf 238(1)}. 1 (2006)



\bibitem[Schlemmer et al.(1994)]{schlemmer1994}
Schlemmer, S., Cook, D., Harrison, J., et al.:  Science {\bf 265}, 1686 (1994)

\bibitem[Schutte et al.(1993)]{sch93} 
Schutte, W.~A., Tielens, A.~G.~G.~M., \& Allamandola, L.~J.: \apj\ {\bf 415}, 397 (1993)

\bibitem[Scott and Duley(1996)]{scott1996}
Scott, A., Duley, W.W.:
%The Decomposition of Hydrogenated Amorphous Carbon: A Connection with Polycyclic Aromatic Hydrocarbon Molecules. Astrophys J 
\apjl\ {\bf 472}, L123 (1996)

\bibitem[Scott et al.(1997)]{scott1997} 
Scott, A. D., Duley, W. W., Jahani, H. R.: \apj\ {\bf 490}, L175 (1997)


\bibitem[Sellgren(1984)]{sellgren1984} 
Sellgren, K.: \apj\ {\bf 277}, 623 (1984)

\bibitem[Sellgren(2001)]{sellgren2001} 
Sellgren, K.: Spectrochimica Acta {\bf 57}, 627 (2001)


\bibitem[Sellgren et al.(2007)]{sel07}  
Sellgren, K., Uchida, K. I., Werner, M. W.: \apj, {\bf 659}, 1338 (2007)

\bibitem[Sellgren et al.(2010)]{sellgren2010} 
Sellgren, K., Werner, M. W., Ingalls, J. G., et al.: \apj\ {\bf 722}, L54 (2010)

\bibitem[Sloan et al.(2014)]{sloan2014}
Sloan, G. C., Lagadec, E., Zijlstra, A. A., et al.:  \apj\ {\bf 791}, 28 (2014)

\bibitem[Smith et al.(2007)]{smith2007}
Smith, J.D.T., et al.: 
%Draine BT, Dale DA, Moustakas J, Kennicutt RC, Jr., Helou G, Armus L, Roussel H, Sheth K, Bendo GJ, Buckalew BA, Calzetti D, Engelbracht CW, Gordon KD, Hollenbach DJ, Li A, Malhotra S, Murphy EJ, Walter F (2007) The Mid-Infrared Spectrum of Star-forming Galaxies: Global Properties of Polycyclic Aromatic Hydrocarbon Emission. 
Astrophys. J. {\bf 656}.  770 (2007)

%\bibitem[Snow(2014)]{snow2014}
%Snow, T.P.: in IAU Symposium 297: the Interstellar Diffuse Bands, eds. J. Cami \& N.L.J. Cox, CUP, p. 3 (2014)

\bibitem[Song et al.(2003)]{Song03}
Song, I.-O., Kerr, T. H., McCombie, J. \& Sarre, P. J.: \mnras\ {\bf 346}, L1 (2003)
	
\bibitem[Spoon et al.(2004)]{spoon2004} 
Spoon, H. W. W., Armus, L., Cami, J., et al.: \apjs\ {\bf 154}, 184 (2004)


\bibitem[Sturm et al.(2000)]{stu00} 
Sturm, E. et al.: \aap\ {\bf 358} 481 (2000)

\bibitem[Tielens(2008)]{tielens2008} 
Tielens, A. G. G. M.: \araa\ {\bf 46}, 289 (2008)



\bibitem[Tokunaga and Bernstein(2021)]{tokunaga2021}
Tokunaga, A.T. and Bernstein, L.S.: \apj\ {\bf 916}, 52 (2021)

%\bibitem[Tokunaga et al.(1991)]{tokunaga1991}
%Tokunaga, A. T., Sellgren, K., Smith, R. G., et al.: \apj\ {\bf 380}, 452 (1991)
	
\bibitem[Uchida et al.(2000)]{uchida2000} 
Uchida, K. I., Sellgren, K., Werner, M. W., Houdashelt, M. L.:  \apj\ {\bf 530}, 817 (2000)

\bibitem[Van Kerckhoven et al.(2000)]{vankerckhoven2000} 
Van Kerckhoven, C., Hony, S., Peeters, E., et al.: \aap\ {\bf 357}, 1013 (2000)

\bibitem[van Diedenhoven et al.(2004)]{van04} 
van Diedenhoven, B., Peeters, E., Van Kerckhoven, C., et al.: \apj\ 611, 928 (2004)



\bibitem[Wada et al.(2009)]{wada2009} 
Wada, S., Mizutani, Y., Narisawa, T., \& Tokunaga, A. T.: \apj\ {\bf 690}, 111 (2009)


\bibitem[Wagner et al.(2000)]{wagner2000}
Wagner, D. R., Kim, H., Saykally, R. J.:  \apj\ {\bf 545}, 854 (2000)


\bibitem[Wickramasinghe and Allen(1980)]{wick1980}
Wickramasinghe, D. T., \& Allen, D. A.: \nat\ {\bf 287}, 518 (1980)

\bibitem[Woolf and Ney(1969)]{woolf1969}
Woolf N.J., Ney E.P.: \apj\ {\bf 155}, L181 (1969)

\bibitem[Wu et al.(2010)]{wh10} 
Wu, Y., Helou, G., Armus, L., et al.: \apj\ {\bf 723}, 895 (2010)

\bibitem[Yamagishi et al.(2012)]{yamagishi2012}
Yamagishi, M., Kaneda, H., Ishihara, D., et al.: \aap\ {\bf 541}, 10 (2012)

%\bibitem[Yang et al.(2017)]{yang2017}
%Yang, X. J., Glaser, R., Li, A., Zhong, J. X.: New Astronomy Reviews {\bf 77}, 1 (2017)

\bibitem[Zagury(2021)]{zagury2021}
Zagury, F.: \aap\ {\bf 652}, L5 (2021a)

%\bibitem[Zagury(2021b)]{zagury2021b}
%Zagury, F.: astro-ph (2021b)

\bibitem[Zhang and Kwok(2011)]{zhang2011}
Zhang, Y., Kwok, S.:\apj\ {\bf 730}, 126 (2011)

\bibitem[Zhang and Kwok(2013)]{zhang2013}
Zhang, Y., Kwok, S.:
%On the detections of C$_{60}$ and derivatives in circumstellar environments. 
Earth, Planets, and Space {\bf 65}, 1069 (2013)

\bibitem[Zhang and Kwok(2015)]{zhang2015}
Zhang, Y., Kwok, S.: \apj\ {\bf 798}, 37 (2015)

\bibitem[Zhang et al.(2010)]{zhang2010} 
Zhang, Y., Kwok, S., Hrivnak, B. J.: \apj\ {\bf 725}, 990 (2010)


\end{thebibliography}
\end{document}